\documentclass[12pt]{article}
\usepackage{epsfig}

\def\beq{\begin{equation}}
\def\eeq#1{\label{#1}\end{equation}}
\def\eeqn{\end{equation}}

\def\beqa{\begin{eqnarray}}
\def\eeqa#1{\label{#1}\end{eqnarray}}
\def\eeqan{\end{eqnarray}}

\let\bar=\overbar

\def\Dslash{\not{\hbox{\kern-4pt $D$}}}
\def\dslash{\not{\hbox{\kern-2pt $\del$}}}

\def\msb{{\bar{\ssstyle M \kern -1pt S}}}

\usepackage{fancyhdr,graphicx}
\def\Title#1{\begin{center} {\Large {\bf #1} } \end{center}}

\begin{document}

\Title{Predictions for signatures of the quark-nova in superluminous supernovae}

\bigskip\bigskip


\begin{raggedright}

{\bf Rachid Ouyed\index{Ouyed, R.}$^{1,2}$, Denis Leahy$^1$, and Prashanth Jaikumar\index{Ouyed, R.}$^{3,4}$}\\
{\it
$^1$Department of Physics and Astronomy, University of Calgary, Calgary, Alberta, Canada; {\it ouyed@phas.ucalgary.ca}\\
$^2$On sabbatical leave  at the Canadian Institute for Theoretical Astrophysics, University of Toronto, Toronto, Ontario, Canada\\
$^3$ Department of Physics and Astronomy, California State University Long Beach, Long Beach, California, USA\\
$^4$ Institute of Mathematical Sciences, CIT Campus, Chennai,
Tamil Nadu, India}
\bigskip\bigskip
\end{raggedright}

\begin{abstract}  Superluminous  Supernovae  (more  than a  100  times
brighter than a typical supernova (SN); e.g.  SN2006gy, SN2005gj, SN2005ap,
SN2008fz,  SN2003ma) have  been  a challenge  to  explain by  standard
models. For  example, pair  instability SNe which  are luminous
enough seem to  have too slow a rise, and  core collapse SNe do
not seem  to be luminous  enough.  We present an  alternative scenario
involving a quark-nova (QN), an explosive transition of the newly born
neutron star  to a  quark star in  which a second  explosion (delayed)
occurs  inside  the  already expanding ejecta  of  a  normal  SN.   The  reheated
SN ejecta  can radiate at  higher levels for longer  periods of
time primarily  due to reduced adiabatic expansion  losses, unlike the
standard SN case. Our model is successfully applied to SN2006gy,
SN2005gj, SN2005ap,  SN2008fz, SN2003ma  with encouraging fits  to the
lightcurves.  There  are four predictions in our  model: (i) superluminous SNe optical 
lightcurves  should show  a double-hump  with  the SN  hump at  weaker
magnitudes occurring days to weeks  before the QN;  (ii) 
Two shock breakouts should be observed vis-a-vis 
one for a normal SN. Depending on the time delay, this would manifest
as two distinct spikes in the X-ray region or a broadening of
the first spike for extremely short delays; 
(iii) The QN deposits
heavy elements of  mass number $A> 130$ at the  base of the preceeding
SN ejecta.  These QN r-processed elements should be clearly visible in
the late spectrum (few days-weeks in case of strong ejecta mixing) of 
the superluminous SN and should be  absent in  the late spectrum  of 
normal  (single explosion) SNe; (iv) The QN yield will 
also contain lighter elements (Hydrogen and Helium). We expect the late spectra 
 to include H$_{\alpha}$
 emission lines that should
 be distinct in their velocity signature  from standard H$_{\alpha}$ emission
  usually attributed to preshocked circumstellar material.  
\end{abstract}

\section{Introduction}

 At high baryon density and vanishing pressure, the ground state of bulk matter may not be iron ($^{56}Fe$), but deconfined strange quark matter (SQM) made up of up, down, and strange (u, d, s) quarks (e.g. Bodmer 1971, Witten 1984)\footnote{Introducing strange quarks into matter costs energy because  the strange quark mass is heavier than that of the up and down quarks. That is why the $\Lambda^0$ baryon (also made up
of an u-quark, a d-quark,  and an s-quark) in vacuum is heavier than the neutron/proton. It decays to them via weak interactions. But as  the baryon density is increased, up and down quarks Fermi levels go higher and higher. One needs 2 d-quarks for every u-quark to maintain neutrality. At some point, this becomes more expensive than introducing the s-quark because the mass of the s-quark is compensated by the fact that one needs only 1up, 1 down and 1 s-quark to make a zero charge unit. So the Fermi level of the d-quark comes down (in other words, the d-quark will decay to an s-quark as density becomes higher and higher). Eventually, we end up with SQM at high density. In general weak decays are prevented in bulk degenerate matter by filled Fermi sea. E.g., the neutron
is unstable in vacuum but relatively more stable in a neutron star (the direct urca does occur
but the lifetime of neutron is much longer).}. In that case, once the density for a transition to the (u, d, s) phase is reached in the core of a neutron star, it is likely that the entire star is contaminated and converted into a (u, d, s) star (e.g. Haensel et al. 1986; Alcock et al. 1986). However, the SQM hypothesis faces skepticism for lack of direct evidence of its existence and theoretical uncertainty regarding the threshold density (see Appendix A for a discussion on SQM and the disaster scenario). 

In 2000, researchers at CERN evaluated results from seven separate experiments (involving collisions between large nuclei) to show that a state of matter in which quarks were not confined had been created (e.g. Margetis et al. 2000). Experiments  at the  relativistic heavy-ion collider (RHIC) at Brookhaven National Laboratory have since conclusively discovered a liquid-like  state of deconfined but interacting quarks and gluons termed the Quark-Gluon Plasma (sQGP) at high
temperature (eg. Nagle \& M\"uller 2006). Certain features of these experiments prove beyond doubt that for a short instant one has created, and in fact significantly exceeded, the conditions required for quark deconfinement. This result, along with the theoretical advance in color superconductivity  in cold and dense matter (eg. Rapp et al. 1998; Alford et al. 1998; Alford et al. 1999) revived the topic of SQM and  formation mechanisms of quark stars and opened doors for the exploration of the post-neutron-star-pre-black-hole stage in the evolution of compact stars.

Ouyed, Dey, \& Dey (2002; hereafter ODD)  considered the intriguing possibility that the transition from a neutron star to a quark star might in fact be an explosive one  and involves two steps in a scenario called the Quark-Nova (QN). First, the neutron star core converts to (u,d) matter. Then, as shown in  Ker\"anen, Ouyed, \& Jaikumar (2005; hereafter KOJ),  the dense (u, d) core collapses to the corresponding stable, more compact (u, d, s) configuration faster than the overlying material (neutron-rich hadronic envelope) can respond. In other words, the growing quark core collapses before it engulfs the overlaying layers of the parent neutron star. While this proposition of an explosive transition awaits more detailed theoretical and numerical investigations in order 
 to be confirmed with certitude, it nevertheless found  many interesting applications in high energy 
  astrophysics  phenomena such as Gamma-ray Bursters (Staff, Ouyed, \& Bagchi 2007; Staff, Niebergal, \& Ouyed 2008),
   Ultra-High Energy Cosmic Rays (Ouyed, Ker\"anen, \& Maalampi  2005), Re-ionization (Ouyed,
   Pudritz, \& Jaikumar 2009), and 
    Cooling of magnetars (Niebergal et al. 2009). These studies  hint to the
     idea that QNe do go off in the universe.

In the QN  model, core deconfinement to (u,d) might occur during or after a SN explosion, so long as the central density of the protoneutron star is high enough to induce phase conversion. Theoretical and numerical work is underway on describing the details of the conversion of ordinary
nuclear matter to quark matter inside neutron stars, in an attempt at improving initial efforts (eg. Olinto  1987;  Drago et al. 2007). This is intimately
connected to recent advances in our understanding of the equation of
state of high-density matter and color superconductivity.  Needless to say, 
the propagation of the burning front
associated with this phase transition is crucial in determining the
energetics of the explosion and ejecta dynamics. 

Staff, Ouyed \& Jaikumar (2006), using state-of-the-art simulations showed that massive neutron stars ($\sim 1.5 M_{\odot}$) born with core densities a few times nuclear density  and spinning at millisecond periods are most likely to reach the quark deconfinement density quickly. These heavy neutron stars (with progenitor mass $ > 25 M_{\odot}$) would experience a QN episode generating a second explosion following the first (i.e. the SN). The delay between the first explosion and the second one varies from a few seconds to a few years depending on the birth parameters (mass, period and magnetic field) of the parent neutron star (see Staff, Ouyed, \& Jaikumar 2006).  Here we describe the application of such a delay 
 for superluminous SNe and briefly discuss the  general implication to  high energy astrophysics.

This proceedings note is organized as follows. In section 2, we describe the essential mechanism and features of the QN. In section 3, we describe in detail the efficacy of the dual-shock QN model in describing the outstanding features of a host of superluminous SNe, with emphasis on the successful interpretation of photometric and spectroscopic results. Section 4 contains an overview of QN dynamics, in particular, the explosive photon fireball that opens up the connection to high-energy astrophysical phenomena as well as r-process nucleosynthesis of the heavy elements in a unified way.

\section{The Quark-nova}

The QN model describes the birth of a quark star (see Appendix A
 on the question of co-existence of neutron stars and quark stars).
In  the QN model,  the {\it (u,d)} core of
a hybrid star\footnote{As  outlined in KOJ, the initial  state for the
QN is that of a deleptonized neutron star with a {\it (u,d)} core.  In
the QN, the two-step process, neutron to {\it (u,d)}, then {\it (u,d)}
to {\it  (u,d,s)} is  crucial.  In our  case, the neutrinos  come from
weak reactions  at the edge of the  {\it (u,d)} core and  can leak out
easily  into the  surrounding cooler  and deleptonized  envelope where
they can  deposit energy.  This is significantly  different from phase
conversion in  a proto-neutron star stage where  neutrino transport is
slower (of  the order of seconds)  because of the  hot and lepton-rich
matter.},  that undergoes the  phase transition  to the  {\it (u,d,s)}
quark phase, shrinks  in a spherically symmetric fashion  to a stable,
more compact  strange matter configuration faster  than the overlaying
material (the neutron-rich hadronic  envelope) can respond, leading to
an effective  core collapse.  The  core of the  neutron star is  a few
kilometers in  radius initially, but shrinks  to 1-2 km  in a collapse
time  of  about  0.1  ms  (Lugones et  al.  1994).  The  gravitational
potential energy  released during this event is  converted partly into
internal  energy (latent  heat)  and partly  into outward  propagating
shock  waves  which  impart   kinetic  energy  to  the  material  that
eventually forms the QN ejecta.

 There are  three previously proposed  mechanisms for ejection  of the
outer layers of the neutron  star (i.e. crust): (i) Unstable baryon to
quark  combustion  leading  to  a shock-driven  ejection  (Horvath  \&
Benvenuto 1988).   More recent  work, assuming realistic  quark matter
equations of state, argues for strong deflagration (Drago et al. 2007)
that  can expel  surface material.   In  these models  up to  $10^{-2}
M_{\odot}$  can   be  ejected.    These  calculations  focus   on  the
microphysics  and  not  on the  effect  of  the  global state  of  the
resulting  quark core  which  collapses prior  to complete  combustion
(KOJ), leading to conversion only of the inner core ($\sim$ 1-2 km) of
the neutron  star; (ii) Neutrino-driven explosion where  the energy is
deposited in  a thin (the  densest) layer at  the bottom of  the crust
above a  gap separating  it from the  collapsing core (KOJ).   For the
neutrino-driven  mechanism,   the  core  bounce   was  neglected,  and
neutrinos emitted  from the  conversion to strange  matter transported
the energy into the outer regions  of the star, leading to heating and
subsequent mass  ejection.  Consequently, mass ejection  is limited to
about  $10^{-5}M_{\odot}$  (corresponding  to  the  crust  mass  below
neutron drip density) for compact  quark cores of size (1-2) km; (iii)
Thermal  fireball driven ejection  which we  consider for  the present
study. The fireball is inherent to the properties of the quark star at
birth.  The  birth temperature was found  to be of the  order of 10-20
MeV since the collapse is adiabatic rather than isothermal (Ouyed, Dey, \& Dey 2002; KOJ).
In this temperature regime, we assume the quark matter is in one of the 
possible superconducting phases, called the Color-Flavor Locked (CFL;
 see Appendix B for an extended definition) 
phase  (Rajagopal\&Wilczek 2001)  where the
photon emissivity dwarfs the neutrino emissivity (Jaikumar, Rapp, \&
Zahed 2002; Vogt, Rapp, \& Ouyed
2004;  Ouyed, Rapp,  \&  Vogt  2005). This is because the
critical temperature of the CFL phase is high $T_c\sim 50$ MeV, so
that even at temperatures of tens of MeV, there is a large amount
of thermal energy in the CFL photons due to in-medium effects. 
We discuss the photon fireball mechanism in detail in Section \ref{sec:fireball}.

\subsection{Dual-shock Quark-Novae}

Before the QN has occurred, one has the progenitor collapse (the SN proper).
 The conversion from NS to QS depends on the NS central density
 at birth. Progenitors 
 with mass $> 25M_{\odot}$ would lead to a  neutron star heavy enough ($\sim
  1.5M_{\odot}$) to experience the transition. 
For low angular momentum progenitors, the combination
 of a high neutron star core density at birth and, most likely, fall-back material
 would drive the proto-neutron star to a  black hole.
 High angular momentum progenitors (collapsars), will delay 
 the formation  of a black hole for three main reasons:
 (a) the progenitor's core 
 tends to shed more mass and angular momentum as it shrinks
 reducing central core mass and fall-back; 
 (b) high spin keeps the core density of the resulting neutron star from crossing the black
  hole formation limit; 
 (c)  high angular momentum in the material around the core reduces the
accretion rate onto the central object. 
All together,  collapsars seem to provide favorable
 conditions for the QN to occur inside them. 
We take the stellar structure of
a Helium Wolf-Rayet star (e.g. Meynet \& Maeder 2003; Heger et al. 2003; Petrovic et al. 2006) to be representative of the progenitor
and consider the low and high-metallicity cases. The main difference is that the
high metallicity star would have an extended envelope. For
example, for a $30M_{\odot}$ star  a rough estimate yields  $\sim 3$-$5R_{\odot}$ 
while in the low metallicity case, the star envelope cuts off sharply at $\sim 1.5$-$2.5R_{\odot}$.

When the iron-rich broken pieces of QN ejecta (see Ouyed\&Leahy 2009) impact this stellar envelope 
 (the already expanding SN ejecta), they undergo
a shock and become heated.  Let us define a certain time delay, $t_{\rm delay}$,  as the time
       elapsed between the SN and the QN explosion. For delays we are concerned with here
 the SN density is high enough that the QN ejecta is  vaporized at impact, effectively
depositing their kinetic energy  at the base of the SN
envelope. The details of this interaction can be found
 in \S 2 in Ouyed et al. (2009).   The resulting QN shock 
  propagating at speed $v_{\rm QN}$
  reaches the outer edge of the SN ejecta (becomes visible to the observer)
  at distance $R_{\rm QN}$ and at time $t_{\rm delay}+ t_{\rm prop.}$ where $t_{\rm prop.}= R_{\rm QN}/v_{\rm QN}$ is the propagation time delay for the QN shock to reach the edge of the SN ejecta;
   {\it which defines the shock breakout in our model which, of
   course, occurs after  the SN breakout shock}.  As we show in the next section
    this interaction between the QN and SN ejecta leads to lightcurves and durations
     observed in superluminous SNe.

\section{Superluminous SNe}

\subsection{Models for light curves} 

SN2006gy was discovered on  18 September 2006 by  Robert Quimby. It  is a
Type IIn SN characterized by the bright peak magnitude $M_R\sim -22$ mag and its long duration (more than 100 times brighter and  significantly longer-lasting  than a typical SN). The fundamental question is how is it possible to power the observed  lightcurve ($> 10^{51}$ erg in radiation) for so long (hundreds of days).

Ofek et al. (2007) argued that the extreme luminosity could have been produced by collision  between a Type Ia SN's ejecta and circumstellar material (CSM). But Smith et al. (2007) point out that in the SN2006gy case one requires an implausibly massive envelope and that no Type Ia features were seen. Smith et al. (2007) instead argue for a pair instability SN (PISN) model with 22 $M_{\odot}$ of Ni$^{56}$ to account for the peak
luminosity. This is orders of magnitudes more than what is produced in an ordinary core-collapse SN, and would occur only for the most massive stars ($> 150 M_{\odot}$)\footnote{Heger \& Woosley (2003) advanced the idea of a pulsational pair-formation SN and argued that the interaction of successive ejection (separated by a few years) would lead to the extreme luminosity. This would require extreme mass-loss rates (a fraction of a solar mass per year) for such a scenario and it is not clear why such an instability will be limited to only two ejections.}. Even if such progenitors can occur in the local universe, PISN models rise extremely slowly compared to SN2006gy (e.g. Figure 6 in Kawabata et al. 2009). Most recently,  Kawabata et al. (2009) performed optical spectroscopy and photometry of SN 2006gy at late time, 400  days after
the explosion. The fits to the lightcurve (see the right panel in their Figure 6) again requires extremes values with kinetic energy of about $6.4\times 10^{52}$ ergs, ejected envelope mass of about $53 M_{\odot}$, and total mass in Nickel of about 15 $M_{\odot}$.

In this context, the QN model provides a natural explanation. The crucial implications and observational consequences of the QN model are: {\bf 1)}  In a QN, the energy input is delayed from the original core collapse explosion, allowing for re-energization of the SN ejecta at larger radius. A typical QN releases up to $\sim 10^{53}$ erg in kinetic energy (Ouyed, Rapp \& Vogt 2005 and references
 therein). The QN ejecta is expelled at ultra-relativistic
 speeds (Ouyed\& Leahy 2008) and re-energizes the preceding SN ejecta as it catches up and collides with it (Ouyed et al. 2009); and {\bf 2)} the resultant shock wave causes the outer layers of the neutron star to decompress, releasing free neutrons and neutron-rich seed nuclei from the crust that can capture these neutrons to form heavy elements via the r-process (Jaikumar et al. 2007). 

 When Leahy \& Ouyed applied the QN model to SN2006gy (with a delay of $t_{\rm delay}= 15$ days between the SN and the QN), it gave an excellent fit  (see  Figure 1 attached; from Leahy \& Ouyed 2008). They also applied the QN model to  two other superluminous SNe: SN2005ap (Quimby et al. 2007)  and SN2005gj (Aldering et al. 2006). As can be seen from the right panel
  in Figure \ref{fig:sn2006gy}, the QN model gives a good  fit  with the delay between the SN and QN as the only parameter ($t_{\rm delay}$= 40 days for SN2005ap and $t_{\rm delay} = 10$ days for SN2005gj). Here we revisit the dual-shock QN in the context of superluminous SNe 
 applying it to more  candidates. We make specific  predictions as described below.

\subsection{Photometry:  {\it The double-hump}}

\begin{figure}[t!]
\begin{center}
\epsfig{file=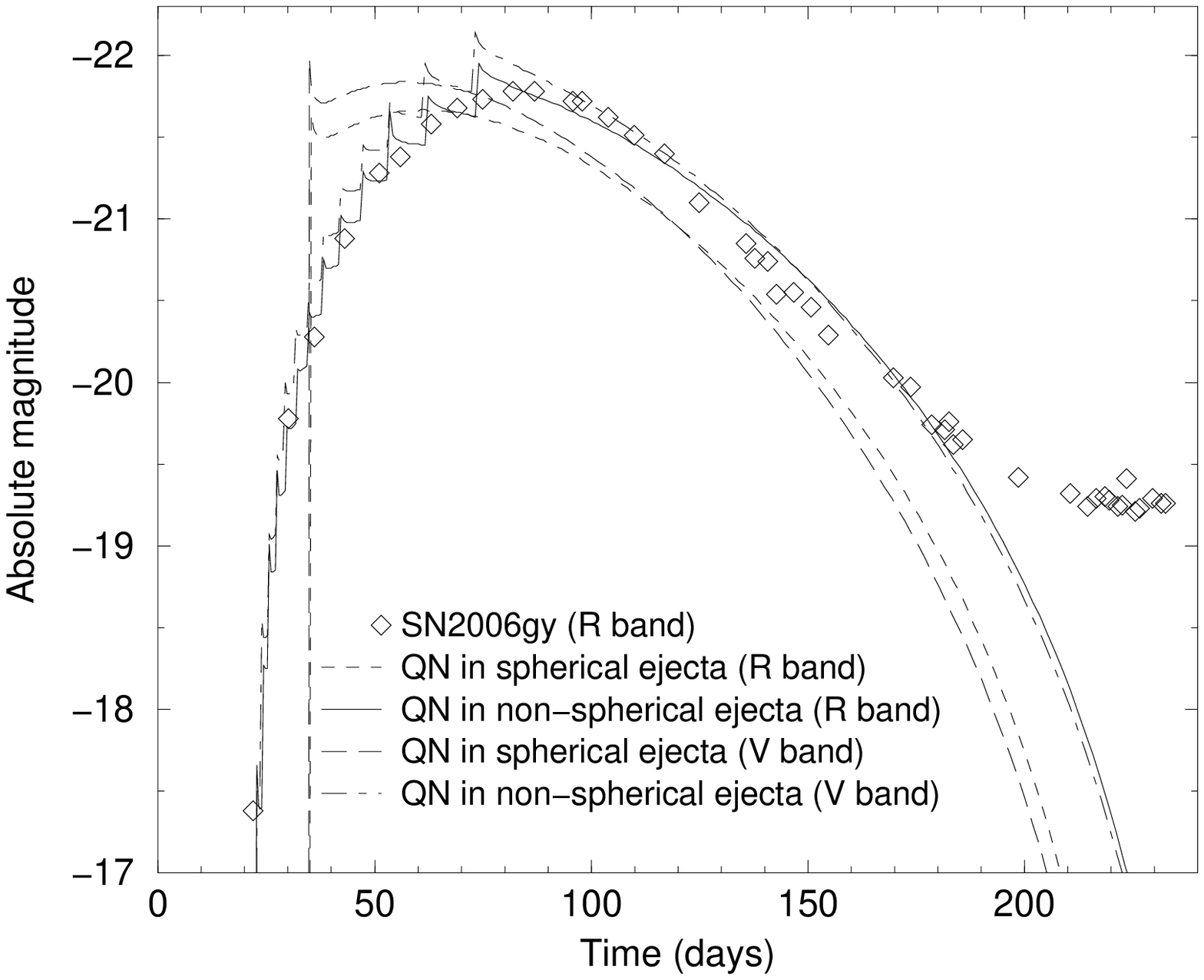,height=2.5in}\epsfig{file=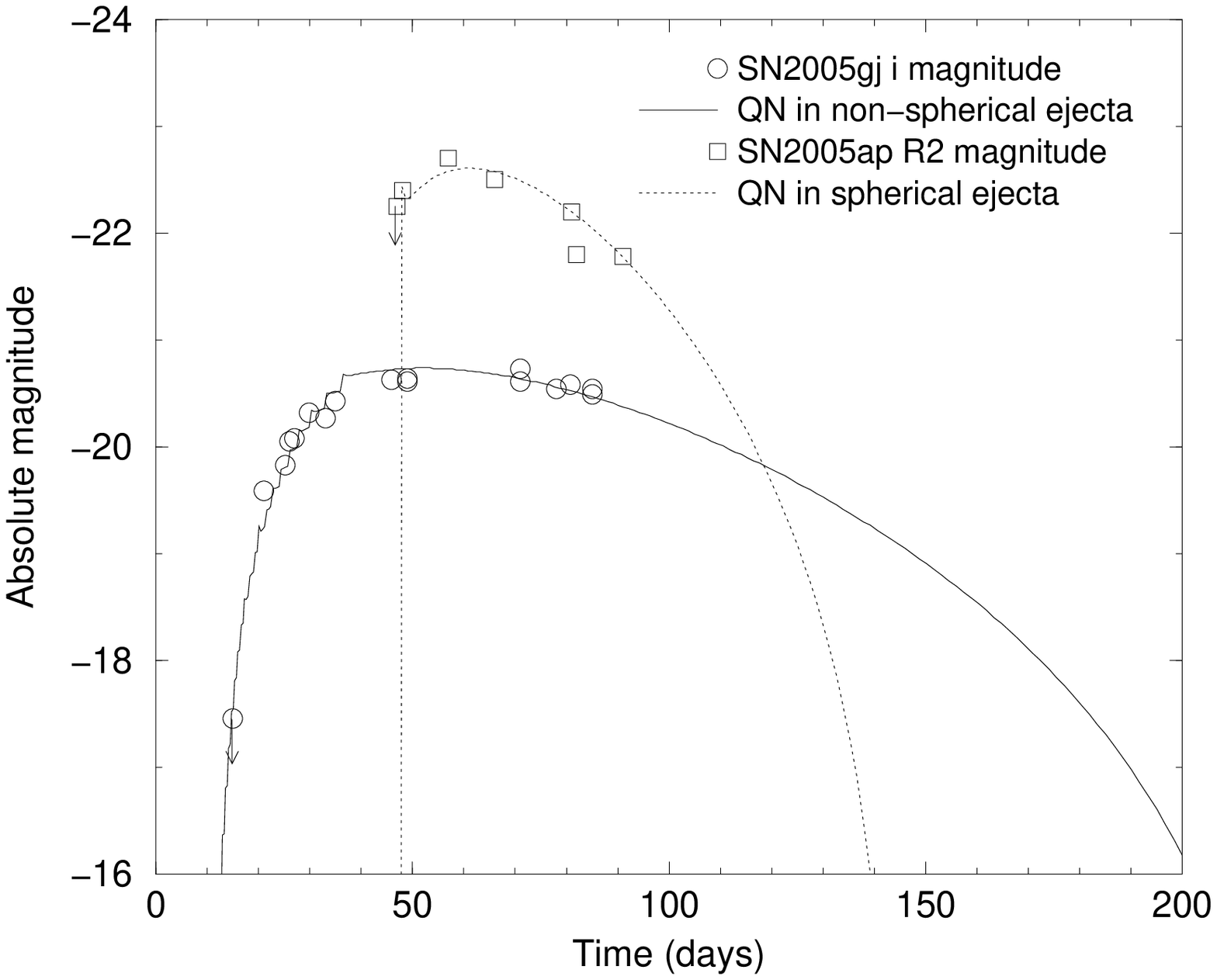,height=2.5in}
\caption{{\bf Left Panel}:  Comparison of the observed  absolute $R$-band light
curve of SN2006gy  and the $R$-band and $V$-band  light curves derived
from our  model (from Leahy  \& Ouyed, 2008).  The  dashed (long-dash)
line  shows the  derived  $R$-band  ($V$-band) light  curve  for a  QN
explosion inside perfectly spherical  SN ejecta.  The solid (dash-dot)
line  shows the  derived  $R$-band  ($V$-band) light  curve  for a  QN
explosion  inside  a  non-spherically  expanding SN  ejecta.   The  SN
parameters  are: Explosion  at  $t=0$, $M_{\rm  eje.}= 60  M_{\odot}$,
$R_0=10 R_{\odot}$,  and the QN  parameters are $t_{\rm  delay}=15$ days,
$v_{\rm QN}  = 6000$ km s$^{-1}$,  and $T_{\rm QN, 0}=  0.4$ MeV.  For
the  spherical case  $v_{\rm  SN}=  3400$ km  s$^{-1}$  while for  the
non-spherical case $2000\ {\rm km s}^{-1} < v_{\rm SN} < 4800\ {\rm km
s}^{-1}$.  The spikes in the derived light curves are due to pieces of
the SN ejecta  being lit up by the QN shock  at different times, which
would  be  smoothed  out   if  the  distribution  of  velocities  were
continuous.  
{\bf Right Panel}: Comparison of the absolute
$i$-band light curve of SN2005gj and $R2$-band light curve of SN2005ap
with  those  derived from  our  model.   For  SN2005gj, the  model  is
calculated with a QN delay of 10  days after the SN, and a range in SN
ejecta speeds  of $750\ {\rm km s}^{-1}  < v_{\rm SN} <  4100\ {\rm km
s}^{-1}$.  For SN2005ap, the model is calculated with a QN delay of 40
days after the SN, and spherical SN ejecta with speed of $v_{\rm SN} =
4000\  {\rm  km  s}^{-1}$.  For  both  models,  all  other QN  and  SN
parameters were kept the same as  for the SN2006gy model. }
\label{fig:sn2006gy}
\end{center}
\end{figure}

 The Leahy \&  Ouyed (2008) model is based  on additional energy input
into the SN ejecta: (i) The explosion occurs inside an extended
expanding envelope;  (ii) The  delay is due  to conversion  of neutron
star (NS) to a quark star  (QS).  We emphasize the crucial role of the
second explosion due to  this delayed conversion.  Benvenuto\& Horvath
(1989)  explored  the  idea  of  conversion energy  release  to  power
SN1987A, which  is a regular SNe.  However, they do  not calculate any
lightcurves or consider explosive conversion making
use of the  conversion delay.  As shown in Leahy  \& Ouyed (2008) this allows
for  more  luminous  and  long-lasting  explosion since  much  of  the
radiation is  emitted rather than  being lost to  adiabatic expansion.
The time-dependent  luminosity of a dual-shock QN  is given by 

\begin{equation}
\label{eq:luminosity} L_{\rm  SN}(t) =  c_{\rm v} \Delta  T_{\rm core}
n_{\rm ejec.}  4\pi R_{\rm phot.}(t)^2 \frac{d D(t)}{d t}\ ,
\end{equation}

\noindent where $c_{\rm  v}\sim (3/2) k_{\rm B}$ is  the specific heat
of  the ejecta,  $\Delta T_{\rm  core}\sim T_{\rm  core}= T_{\rm QN,0} R_0^2/(R_0+ v_{\rm SN} t)^2)$ is  the core
temperature of the ejecta;  $R_0$ is the radius of the progenitor,
 $v_{\rm SN}$ the speed of the SN ejecta, and $T_{\rm QN,0}$
  is the temperature of the SN ejecta when it is first reheated by the QN shock.
Also,  $n_{\rm ejec}$ is the number density of the
ejecta, $R_{\rm  phot.}(t)$ is the  photospheric radius and  $D(t)$ is
the photon  diffusion length (see Leahy\&Ouyed 2008 for mode details). 
 The resulting  light curve  is a superposition  of light  curves from
different parts of the reshocked  SN shell, with different rise times,
different peaks,  and different shapes. The  corresponding light curve in the case
 of SN2006gy 
is shown  in Figure \ref{fig:sn2006gy}  (solid and dash-dot  lines for
$R$- and $V$-band, respectively) and corresponds to an SN explosion at
$t=0$ with  ejecta mass $M_{\rm ejec.}= 60 M_{\odot}$,  $R_0=10 R_{\odot}$, $t_{\rm
delay}=15$ days,  $v_{\rm QN}  = 6000$ km  s$^{-1}$, and $T_{\rm  QN, 0}=
0.4$  MeV.  The  lightcurve was  computed by  averaging over  13 equal
solid angle  segments of a  sphere with different  velocities linearly
spaced between the minimum and maximum values: $2000\ {\rm km\ s}^{-1}
< v_{\rm SN} < 4800\ {\rm km\ s}^{-1}$.  The lightcurve first turns on
when the slowest ejecta ($v_{\rm  SN, min}=2000$ km s$^{-1}$) is fully
reshocked   at    $t_{\rm   delay}+t_{\rm   prop.}(v_{\rm    SN,   min})=
(15+7.5)=22.5$ days.   The Smith et  al. (2007) data was  plotted with
the first  data point  (an upper limit)  at $t=  22$ days in  order to
match our model  with the overall rise.  The  spikes in the lightcurve
(dashed line) are due  to pieces of the SN ejecta being  lit up by the
QN  shock at  different  times, which  would  be smoothed  out if  the
distribution of velocities were  continuous.  The SN material at lower
velocities  experiences  the  QN  shock earlier  resulting  in  larger
adiabatic losses  and lower peak  brightness.  

\begin{figure}[t!]
\begin{center}
\epsfig{file=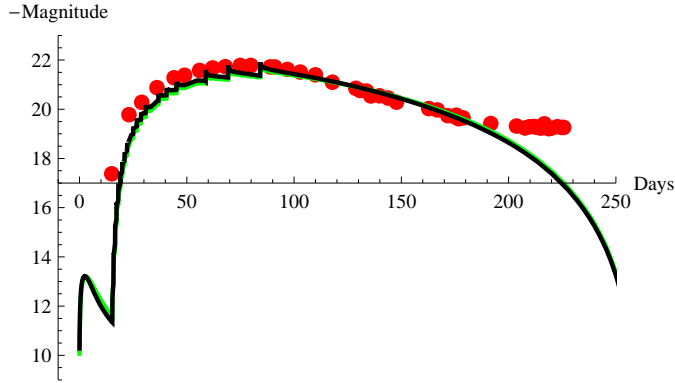,height=2.0in}
\caption{The dual-shock  QN application to SN2006gy. The black (green)
curve is the R-band (V-band) magnitude calculated from the QN model.  
Shown  is the SN  hump preceding
the QN one with a delay  of 15 days between the two explosions. One of
the predictions  of the  QN is these  double hump feature  that should
accompany extremely luminous SNe. Adding the maximum expected
 amount of $^{56}$Ni for  the  $60M_{\odot}$ progenitor 
       will increase the magnitude of the SN hump
        to no more than $\sim -18$.}
\label{fig:doublehump}
\end{center}
\end{figure}

\begin{figure}[th!]
\begin{center}
\epsfig{file=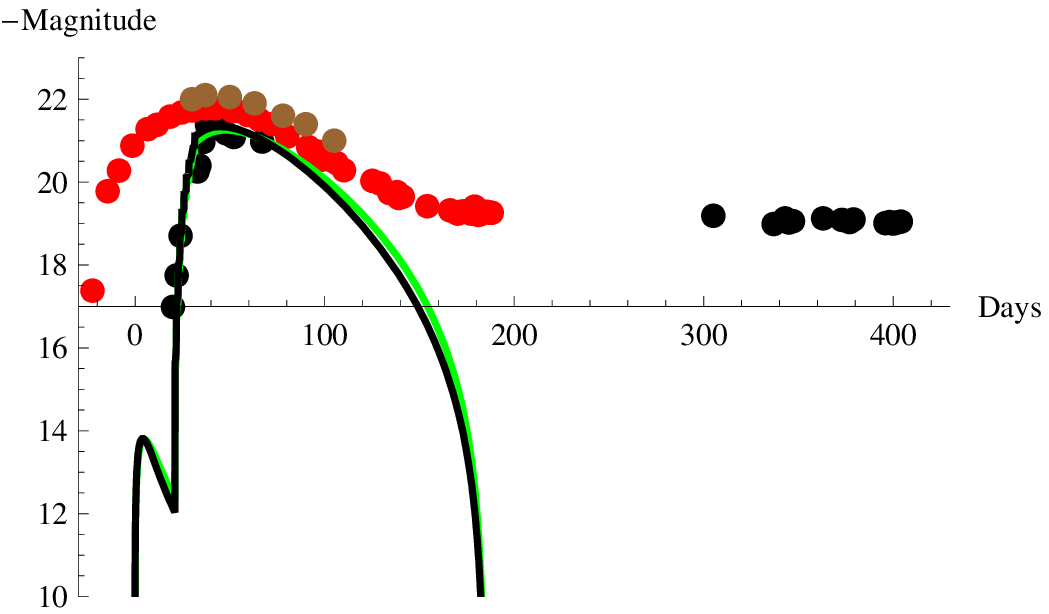,height=1.8in}\epsfig{file=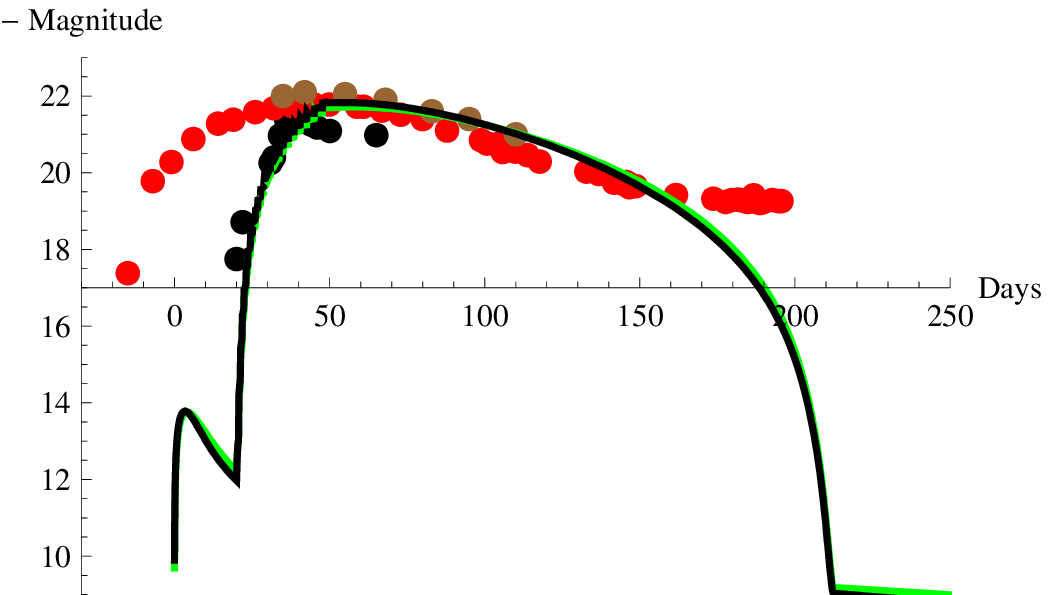,height=1.8in}
\caption{Observed SN2003ma (top-most black dots) and SN2008fz (bottom-most brown dots).
The broader SN2006gy (the red dots in-between) is shown for comparison. 
 Also shown is the predicted SN  hump preceding
the QN for SN2003ma (in the left panel) and SN2008fz (in the right panel). The left
 panel shows the late plateau of SN2003ma which we discuss
  in an upcoming paper (see  \S 3.2 in Leahy\&Ouyed (2008) for a discussion of the plateau
   in SN2006gy). Nickel decay 
       will increase the magnitudes of the SN humps
        to no more than $\sim -18$.}
\label{fig:sn2003masn2008fz}
\end{center}
\end{figure}

In Figures \ref{fig:doublehump} and \ref{fig:sn2003masn2008fz} we show
 the double-humped feature predicted by the QN model
  for SN2006gy, SN2003ma (Rest et al. 2009) and SN2008fz (Drake et al. 2009).  
   We note that  the first
shock (namely the SN proper) might be  too  faint to be seen due to the large
distance to these SNe.  
   We did not include the $^{56}$Ni
    decay contribution. Adding a few solar masses of
      $^{56}$Ni (i.e. the maximum expected for a $60M_{\odot}$ progenitor; Nomoto et al. 2007)
       will increase the magnitudes of the SN humps
        to no more than $\sim -18$ (which might be slightly above the
upper limit  for detection).
       Table 1 summarize the model parameters  fits for the five superluminous
   studied here.  For the chosen SN progenitor ($M_{\rm ejec.}=60M_{\odot}$,
$R_{*}=10R_{\odot}$,  and $T_{\rm SN,0}=0.3$ MeV), observed
superluminous  SNe can be well fitted with  delays times
 between the two explosion in the range: $ {\rm days} < t_{\rm delay} < {\rm weeks}$.


\subsection{Spectroscopy:  {\it The unusual lines}}

Kawabata et al. (2009) found  unusual features in the innermost ejecta
visible at $t=394$ days for  SN2006gy (i.e. in the late time spectrum;
marked with dashed lines in their  Figure 5). These lines are not seen
in other types of SNe (see  their Fig. 9-11), rather resembling Ti and
Ca lines. One of the interesting discoveries in Jaikumar et al. (2007)
was the smaller peaks appearing around mass numbers $A\sim 44$ (mainly
from heavy Ca and Ti isotopes -  see Figure \ref{fig:rprocess}) and $A \sim 80$ (mainly
from Se and Kr and their isotopes) along with production of heavy 
elements up to the 3rd r-process peak. Such a feature follows from
the large neutron-to-seed ratio and strong fission cycling that
a QN can provide naturally.

Simulations of r-process nucleosynthesis  in the QN model (Jaikumar et
al.  2007; Charignon, Ouyed, \& Jaikumar  2009) indicate that the elements near the
3rd r-process peak  are very efficiently produced when  the energy and
entropy level  are those expected  in a QN scenario.   For not
too  small $Y_e\sim  0.12$, these  simulations support  a  strong link
between the  QN and small  peaks appearing around mass  numbers $A\sim
44$ (mainly from heavy Ca, Ar  and Ti isotopes) and $A\sim 80$ (mainly
from Se and Kr and their isotopes; see Figure \ref{fig:rprocess}).

The QN  ejecta is  expected to achieve  gamma-ray transparency
sooner  than  SN  ejecta  since QN  progenitors  (i.e.,
neutron   stars)   lack   extended   atmospheres.  This   means   that
r-process-only nuclei with gamma-decay lifetimes of the order of years
(such  as $^{137}$Cs,  $^{144}$Ce, $^{155}$Eu  and $^{194}$Os)  can be
used as  tags for  a quark star  formed in  a recent quark  nova.  The
gamma-fluxes  from some long-lived  radionuclides such  as $^{226}$Ra,
$^{229}$Th  and $^{227}$Ac  produced in  a QN  have been  tabulated in
Jaikumar et al.  (2007). These results provide an  observable that can
distinguish a QN  from a SN, and they  also establish a
connection  to  the  r-process  that  can  be  empirically  tested  by
satellite-based instruments in the very near future.

\begin{table}[t!]
\begin{center}
\begin{tabular}{|c|c|c|c|c|c|}\hline
SN  &   $t_{\rm delay}$ (days)  &   $T_{\rm QN, 0}$ (MeV) & $v_{\rm QN,
shock}$ (km/s) & $v_{\rm SN, min}$ (km/s)  &  $v_{\rm SN, max}$ (km/s) \\
\hline
2006gy   &  10  & 0.4 & 6000 &  2000 & 4800 \\ \hline
2005ap   &  40  & 0.4 & 25000 &  4000 & 4000 \\ \hline
2005gj   &  10 & 0.4 & 6000 &  750 & 4100 \\ \hline
2008fz   &  20  & 0.9 & 8000 &  200 & 4800 \\ \hline
2003ma   &  30  & 0.5 & 8000 &  2900 & 2900 \\ \hline
\end{tabular}
\caption{Dual-shock QNe fits to superluminous SNe. 
 Shown are the delay time between the two explosion ($t_{\rm QN}$),
  the initial temperature of the reshocked SN ejecta ($T_{\rm QN,0}$),
   the shock velocity of the reshocked SN ejecta  ($v_{\rm QN,shock}$)
    and the  range in the SN ejecta's velocities ($v_{\rm SN, min}$
     and $v_{\rm SN,max}$).
The underlying SN parameters,
 kept fixed for all candidates, are $M_{\rm ejec.}=60M_{\odot}$,
$R_{*}=10R_{\odot}$,  and $T_{\rm SN,0}=0.3$ MeV.  
Candidates with  $v_{\rm SN, min}\sim v_{\rm SN, max}$ were best fit with an almost
spherical shell  (asphericity not exceeding the 10\%).}
\end{center}
\label{tab:all}
\end{table}

\section{Discussion and Conclusion}

\subsection{The photon fireball and explosive astrophysics}
\label{sec:fireball}

In core-collapse SNe, neutrinos being the lightest and most weakly interacting particle, carry away 99\% of the star's binding energy  and drive the explosion. In QNe, neutrinos emitted from the quark core have long diffusion timescales, of order 10-100 ms for $T\sim 20$ MeV (KOJ) and cannot escape before the entire star converts to $(u,d,s)$ matter. With a hadronic crust, the heating effect of the neutrinos implies that the mean free path in the crust is of the order of meters, and mass ejection is of the order of $10^{-5}M_{\odot}$. Thus, neutrinos are not very efficient in depositing their energy in the outer layers of the star in a short enough time. 

For a QN, the suitable agent of explosion are the photons, since the temperature of the quark core is large enough at the time of formation ($\sim 20-50$ MeV) to sustain large photon emissivities. If quark matter is in a normal (i.e., non-superconducting) state ($T\geq 50$ MeV), the photon emissivities are extremely large since $T>\hbar\omega_p/k_B$ where $w_p\sim (20-25)$ MeV is the plasma frequency of normal quark matter (Usov 2001). The mean free path is small enough to thermalize these photons inside quark matter, and smaller still in the hadronic envelope ($\omega_p({\rm hadronic})\sim 100$ MeV) so that energy deposition by photons is highly efficient. 

If quark matter is superconducting and in the CFL phase, the result is similar. A huge build-up of thermal photons occurs because of medium effects. At $T=0$, a residual but exact $U(1)$ symmetry ensures that CFL photons travel freely in the quark medium. However, at $T\sim (5-50)$ MeV, these photons thermalize due to electromagnetic interactions with charged light Goldstone mesons (Vogt, Rapp \& Ouyed 2004; Jaikumar, Prakash \& Sch\"afer 2002) and their mean free path is of order fermis. The resulting emissivity saturates the black body limit at $T\sim 50$ MeV. Compared to the neutrino flux from hot quark matter, the photon flux is from 1-3 orders of magnitude higher
for temperatures in MeV-tens of MeV. It follows that energy deposition in the crust is much more efficient for photons than neutrinos. Even a few percent of the photon energy, when deposited in the thin crust of the star, will impart a large momentum to it, leading to strong and ultra-relativistic mass ejection (Ouyed\&Leahy 2008). Up to $10^{-2}M_{\odot}$ can be ejected by the photon fireball, making the QN a highly explosive and luminous astrophysical phenomenon. As we have argued on the basis of observations as well, this lends itself naturally to an explanation of the superluminous SNe. Below, we discuss some astrophysical consequences and predictions that follow from the QN.

\begin{figure}[t!]
\begin{center}
\epsfig{file=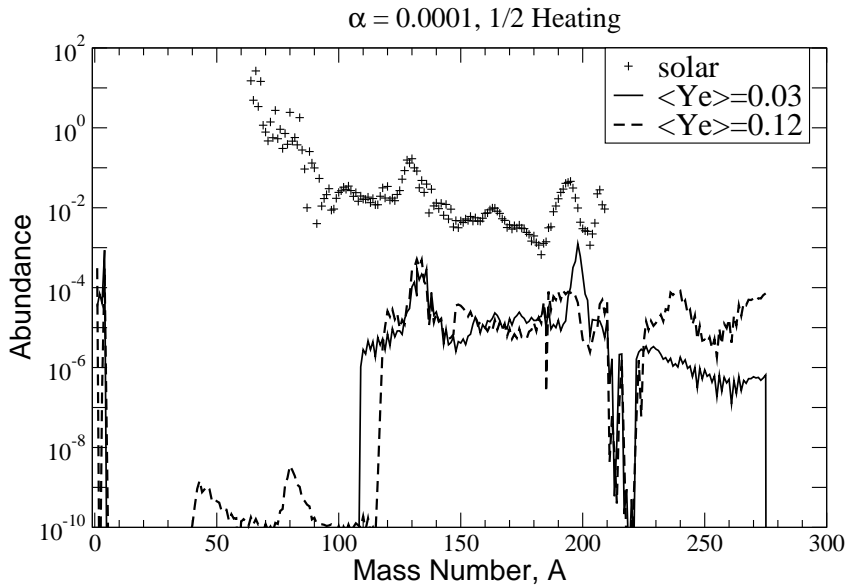,height=5.0in,angle=-90.0,}
\caption{The r-process element abundance distribution as a function of 
mass number in a QN: parameters of the simulation are fixed
to expansion timescale $\tau\sim 0.1$ms, 50\% local energy deposition from
 nuclear reactions, and mass averaged electron fraction of ejecta 
$\langle Y_e\rangle$=0.03 and 0.12 as shown (from Jaikumar et al. (2007)).
 Note the peaks at Hydrogen and Helium.
 The solar heavy-element abundance is also shown for comparison.}
\label{fig:rprocess}
\end{center}
\end{figure}

\subsection{The QCD-Astrophysics connection}

The discovery of a QN would have tremendous implications for
Quark Matter and astrophysics. Among the immediate implications:

\begin{itemize}

\item It would  confirm that SQM (in a color superconducting, CFL-like,
 state) exists in a stable state in the universe. 

\item The  separation between the humps
 in the lightcurve (i.e. $t_{\rm delay}+t_{\rm prop.}$) is an estimate of the time delay
   between the two explosions, $t_{\rm delay}$.  This delay is intimately linked
   to the value of the deconfinement density   (see Staff, Ouyed, \& Jaikumar 2006)
    which is of relevance to QCD.
  
 \item The additional energy deposited by the QN ejecta is an estimate of the
  energy release during a QN explosion. Since the gravitational energy 
is transferred to internal and shock energy, superluminous SNe
provide a window into the energetics of the phase transition in
cool and dense matter.

\item QN do not require the existence of extreme mass progenitors (see Appendix C). If the proposed observations reveal the expected sequence of spectroscopic features, our views  of the final states of massive stars have to be revised. Even a small fraction of progenitors with masses in the 25-40 $M_{\odot}$ range undergoing a double explosion is enough to account for the rate of observed superluminous SNe.

\item The first results on the production of heavy elements (with $A > 130$) in a QN described in Jaikumar et al. (2007; see also Charignon et al. 2009) were meaningfully compared to the striking r-process pattern seen in several old metal-poor stars. The observed delamination of a QN would make this connection more robust with immediate implications to our modeling of chemical evolution of galaxies and of the
IGM. As shown in Ouyed, Pudritz, \& Jaikumar (2009), with  QNe, a normal initial mass function (IMF) for the oldest stars can be reconciled with the mean metallicity of the early IGM post-reionization. It would be remarkable if the solution to the longstanding problem of heavy-element ($A >130$) nucleosynthesis  and the early IGM metallicity  enrichment is intimately linked to the conclusive spectral evolution of a dual-shock QN.

\item Yasutake et  al. (2005) and Staff  et al. (2006)  have determined that
the  evolutionary transition  from rapidly  rotating neutron  stars to
quark  stars  due   to  spin-down  can  lead  to   an  event  rate  of
$10^{-4}$-$10^{-6}$ per  year per  galaxy. Similar rates  were derived
from studies of QNe contributions  to r-process material in the Galaxy
by Jaikumar et al. (2007) who  estimated that 1 out every 1000 neutron
stars might  have undergone  a QN.  Since  the Galaxy  likely contains
about  $10^{8}$ neutron  stars this  suggests  an average  QN rate  of
$10^{-5}$  per year  per galaxy.   Interestingly, the  fraction  of SN
progenitors with  mass greater than $60M_{\odot}$ can  be estimated as
$\sim 5\times  10^{-3}$, using the Scalo (1986)  initial mass function
for $M> 8 M_{\odot}$.  Using a  SN rate of $\sim 10^{-2}$ per year per
galaxy,  we get $\sim  5\times 10^{-5}$  per year  per galaxy  for the
explosion rate  of massive  star ($> 60  M_{\odot}$). This  is, within
uncertainties, the same as the QN rate.

\end{itemize}

\subsection{Four predictions}

\begin{itemize}

\item Two SN-like lightcurves are seen, with  the first a normal SN lightcurve,
and  the second  brighter  one delayed  from  the first  by  by a  few
days to a few weeks. The shape and peak  magnitude of the second lightcurve depend on
the delay time (with generally lower peak for shorter delays) and also
on the SN envelope mass and SN ejection velocity.

\item Two shock breakouts should be observed for the QN vis-a-vis 
one for a normal SN. Depending on the time delay, this would manifest
as two distinct spikes in the X-ray region or a broadening of
the first spike for very short delays ($< 1$ day) between the two explosions. 
 The double shock breakout 
 specific to the QN model is currently being investigated in more details.

\item Another tell-tale signature that would clearly
distinguish a QN from a SN would be the detection of radioactive elements with
lifetimes of order several days or few weeks. Such short-lived elements
cannot realistically be detected in SN, since their decay times are too
short compared to the gamma transparency of the SN ejecta. But the QN
ejecta is relativistic and plows into the SN ejecta at very large speeds,
producing the dual shock. If subsequent mixing effects are strong, there is 
the possibility that some of the short-lived heavy elements could be detected 
against the background of the late-time (decaying) light curve. Based on
the r-process calculations of Jaikumar et al. (2007), we have identified 
some promising ``signature'' elements of the dual QN, including but not 
limited to:$^{156}$Eu$_{63}$, $^{166}$Dy$_{66}$, $^{175}$Yb$_{70}$,
$^{183}$Ta$_{73}$, $^{191}$Os$_{76}$, $^{193}$Ir$_{77}$,
$^{223}$Ra$_{88}$, $^{225}$Ac$_{89}$. (All of them are 
$\gamma$-active with lifetimes of few days-few weeks): 
Specifically, gamma decay lines from these elements
should be seen in late  spectra of superluminous
SNe once the photosphere has receded deep into the ejecta.

\item As seen from Figure \ref{fig:rprocess}, the QN yield will 
also contain hydrogen and helium. We expect the spectra in dual-shock QNe (superluminous SNe)
 to include  H$_{\alpha}$
 emission lines  that should
 be distinct (delayed) from standard H$_{\alpha}$ emission
  usually attributed to preshocked circumstellar material. 
  There are a few caveats attached to this fourth prediction: (i)  As can be seen
 from Figure \ref{fig:rprocess}, the hydrogen (and helium) abundance
  is no higher than the heavy element peaks. 
   Since H and He are two elements whereas the band from A=100 to 250 has 150 elements, 
 H would comprise only 1\% by number (i.e. $ \sim 10^{-4}$ by mass).
  Thus there will be no more than   $\sim 10^{-4}M_{\odot}$ of hydrogen in the QN ejecta.
   Once mixed with the preceding SN material the H might be hard to detect
    despite its   distinctive  velocity signature (with line-width $\propto v_{\rm QN}$);
 (ii)  If  the H is moving fast outward in a spherical shell, the line-of-sight
   Doppler velocity will range from $-v_{\rm QN}$ to $+v_{\rm QN}$, 
   so that the line is spread over a wide frequency range and thus be very faint;
  (iii) if there is too much H$_{\alpha}$ emission in the SN associated galaxy, the
    QN H$_{\alpha}$ emission line  would not be detectable, 
 even if it is at high velocity.

\end{itemize}

Finally, it should be emphasized that there  are several active research  
fronts on understanding the possible implications of 
 the  QN to other outstanding astrophysical phenomena
(e.g.   gamma-ray
 bursters, ultra-high energy cosmic rays, re-ionization ect ...).
 Some of these  phenomena have been listed in the  ``Turner report"
    which was entitled ``{\it Connecting Quarks with the Cosmos: Eleven Science Questions for the New Century}"
    (Committee On The Physics Of The Universe, 2003). 
It  would  be   remarkable  for  example  if  the   solution  to  some of these 
longstanding  problems  could find  answers in the discovery of
stable quark matter in the universe via a QN. A Quark-Nova is a phenomenon that would naturally connect  quarks with the cosmos; the most powerful explosions in the universe
 signaling the birth of the tiniest of particles.

\bigskip
R.O. is grateful to  organizers of the CSQCDII conference. R. O. thanks
 the Canadian Institute for Theoretical Astrophysics (CITA) for  hospitality and 
  support  during the completion of this work.  The research of R. O. and
   D. L. is supported by an operating grant from the National Science and Engineering
    Research Council of Canada (NSERC).

\begin{appendix}

\section{Disaster scenarios and pollution of neutron stars by CFL
strangelets}

Here, we would  like to discuss ``myths" related to strange matter
and the destruction of the hadronic universe, and  the co-existence of 
neutron and quark stars:

(i)  One might argue that if a strangelet comes in contact with a lump
   of ordinary matter (e.g. the Earth), it could convert ordinary
    matter to strange matter thus leading to the so-called "ice-nine"
disaster scenario.  Madsen (2001) showed that the reduction of the number of
strange quarks near the surface of a CFL strangelet is energetically
preferable, making it positively charged and electrostatically
repelled by nuclei. In the interior of a
neutron star, neutrons are readily converted since they  do
not repel the seed of strangelets and, to a first
approximation, would instead contribute to the growth of the strangelet cloud 
eventually converting the entire star.

(ii)  The point above implies that pollution by strangelets
(e.g. ejected during coalescence of quark stars in binary systems; e.g.
Madsen 2005) would have
 converted all of neutron stars to quark stars.
However one can show that because of the stiffness of quark matter compared
 to neutron matter, the flux of ejected CFL strangelets during binary
coalescence is very low and is below
 current detection limits (see discussion in Appendix A in Ouyed, Pudritz, \& Jaikumar , 
2009). Thus, at present, the strange quark matter hypothesis is not
inconsistent with observations. The most recent work about strange star 
binary mergers by Bauswein et. al. (2008)  found that for large values of the 
MIT bag constant, strange stars can co-exist with ordinary neutron stars as 
they are not converted by the capture of cosmic ray CFL strangelets. Combining 
their simulations  with recent estimates of stellar binary populations,
Bauswein et al. (2008)  conclude that an unambiguous detection of an
ordinary neutron star would  not rule out the strange matter
hypothesis.

\section{Color superconductivity: {\it The Color-Flavor Locked phase}}

In theoretical terms, a color superconducting phase is a state in which the quarks near the Fermi surface become correlated in Cooper pairs, which condense. The dominant interaction between quarks is the strong interaction, described by QCD, which is  attractive in some channels (the same
 force that binds quarks together to form baryons). It is thus  expected that quarks will form Cooper pairs very readily and that quark matter will generically acquire a condensate of Cooper pairs (Ivanenko\&  Kurdgelaidze, 1969;  Barrois 1977;  Bailin\& Love, 1984).

 Since there are  three different colors (red, green, blue) , and three different flavors (up, down, and strange)  color superconducting quark matter can come in a rich multiplicity of different possible phases, based on different pairing patterns of the quarks. Thus in forming the Cooper pairs there is a 9 by 9 color-flavor
matrix of possibilities in accordance with Fermi statistics (this means the number becomes much less than 81 once the antisymmtery condition is imposed). The differences between these patterns are very physically significant: different patterns break different symmetries of the underlying theory, leading to different excitation spectra and different transport properties. Which type of pairing is favored at which density remains to be answered.  Only at the
highest densities can perturbative calculations be applied confidently (Rajagopal\&Wilczek 2001).   For such extreme densities, the mass of the strange quark is negligible compared to
    the baryonic chemical potential, leading to the same density of the three flavors u,
    d and s quarks. Consequently, it also implies that the CFL phase is naturally electrically
     neutral.  In this special regime, it is found that the  color-flavor-locked (CFL) quark pairing, in which all three flavors participate symmetrically (with total zero momentum), is favored. 
     The CFL phase is so named because the condensate is not separately invariant under color and flavor transformation - it is only under a combined transformation of color and flavor (a "locking")  that it remains invariant. 
     CFL quark matter has many special properties, including the fact that chiral symmetry is broken by a new mechanism: the quark pairs themselves, instead of the more conventional chiral condensate. These 
      properties of the CFL phase have interesting implications to astrophysics
       such as the generation of a photon fireball (see discussion in \S \ref{sec:fireball}) which was shown to be of relevance
        to gamma-ray bursters (Ouyed, Rapp, \& Vogt, 2005).

\subsection{The Meissner Effect}

In superconducting metals, 
the condensate of Cooper pairs of electrons is charged (breaking the charge symmetry), and as a result the photon, which couples to electric charge, becomes massive. Superconducting metals therefore contain neither electric nor magnetic fields (the Meissner effect). In color superconductivity, 
 since pairs of quarks cannot be color-neutral, the resulting condensate will break the local color symmetry, making the gluons massive.  As the photon enters the  superconducting quark, it sees the electric charge of the quark. The quark charge is  related to flavor via the vector part of the flavor symmetry, but since the flavor symmetry is locked to color in the CFL phase, so the photon will indirectly also see the  color charge of the quark. This is equivalent to photon-gluon mixing.   The photon itself does not become massive, but mixes with one of the gluons (the eight) to yield a new massless ``rotated photon" which has no Meissner effect living happily inside the CFL (Alford et al. 2000). 
 
 The conclusion above led to the believe that the magnetic field expulsion in CFL quark stars
  is unlikely to occur. However this might not be totally true. For example, 
  while most color superconductors (with Spin-0 pairing) are not electromagnetic superconductors
  because of the ``rotated electromagnetism",  it is not the case
   in Spin-1 color superconductors where the Meissner effect is real (e.g.
    Schmitt et al. 2003).  There are other reasons one should expect magnetic
     field expulsion in CFL quark stars.
 In a rotating CFL quark star  (the QN compact remnant)
rotational vortices  are also formed (since the CFL phase is a superfluid). 
 The attractive force between the magnetic field confined to the 
   rotational vortices  and the one in the bulk matter
    would lead to expulsion of the bulk magnetic field as the
     star  spins-down and expels vortices. This is equivalent
      to an effective Meissner-like effect with interesting implications
       to cooling of compact stars  (see Niebergal et al. 2009 for
        more details).

\section{Progenitors of dual-shock Quark-Novae}

The fit to the observed light curve of the superluminous SNe studied here  assume
 QNe progenitor mass in the (40-60)$M_{\odot}$ range.
 However, one can employ the parameter degeneracy in that fit to examine the dual-shock scenario with the more conservative mass range of (25-40)$M_{\odot}$ which is more in line with the literature  (e.g. Heger et al. 2003; Nakazato et al. 2008) which suggests prompt BH formation above 
 $40M_{\odot}$.  According to Ouyed et al. (2009), the shock efficiency varies 
  with the SN envelope  density as as $\rho_{\rm env.}^2$ with the mean
 SN envelope density  given by $\rho_{\rm env}\propto M_{\rm env.}/R_{\rm env}^3$. If we choose
30$M_{\odot}$ instead of 60$M_{\odot}$ for the progenitor of the SN,
and demand the same efficiency, we find that
the collision radius should be $R_{30} =
(30/60)^{1/3}\times R_{60}\sim 0.8\times R_{60}$ and the delay time $t_{\rm delay, 30} = 0.8 t_{\rm delay, 60}$; i.e. time delays  shorter by 20\%.
  It should be noted however that the effect of the fireball in the CFL phase 
 (Ouyed, Rapp, \& Vogt, 2005) has not been taken into account in any of these simulations, which means the range 25-40$M_{\odot}$ could be somewhat underestimated.

\end{appendix}

\end{document}